\documentclass[twocolumn,english,preprintnumbers]{revtex4}
\usepackage{amssymb}
\usepackage{amsmath}
\usepackage{colordvi}
\usepackage[colorlinks]{hyperref}
\usepackage{amsthm}
\usepackage{subeqnarray}
\usepackage{cases}
\usepackage{enumerate}
\usepackage{graphicx}
\usepackage{epsfig}
\usepackage{epstopdf}
\usepackage{subfigure}
\usepackage{color}
\usepackage{url}
\usepackage{verbatim}
\usepackage{mathrsfs}

\usepackage{color}

\begin{document}

\title{Self-bound droplet clusters in laser-driven Bose-Einstein condensates}
\author{Yong-Chang Zhang$^{1}$}
\email{zhyongchang@hotmail.com}
\author{Valentin Walther$^{1,2}$}
\author{Thomas Pohl$^1$}
\affiliation{$^1$Center for Complex Quantum Systems, Department of Physics and Astronomy, Aarhus University, Ny Munkegade 120, 8000 Aarhus C, Denmark\\$^2$ITAMP, Harvard-Smithsonian Center for Astrophysics, Cambridge, Massachusetts 02138, USA}

\begin{abstract}
We investigate a two-dimensional Bose-Einstein condensate that is optically driven via a retro-reflecting mirror, forming a single optical feedback loop. This induces a peculiar type of long-range atomic interaction with highly oscillatory behavior, and we show here how the sign of the underlying interaction potential can be controlled by additional optical elements and external fields. This additional tunability enriches the behavior of the system substantially, and gives rise to a surprising range of new ground states of the condensate. In particular, we find the emergence of self-bound crystals of quantum droplets with various lattice structures, from simple and familiar triangular arrays to complex superlattice structures and crystals with entirely broken rotational symmetry. This includes mesoscopic clusters composed of small numbers of quantum droplets as well as extended crystalline structures. Importantly, such ordered states are entirely self-bound and stable without any external in-plane confinement, having no counterpart to other quantum-gas settings with long-range atomic interactions. 
\end{abstract}

\maketitle

\section{Introduction }
Self-organization, the phenomenon that particles spontaneously order in regular patterns or other stable structures, is a ubiquitous phenomenon in nature that has been attracting broad scientific interest for a long time \cite{Cross}. This behavior is closely connected to spontaneous symmetry breaking and the occurrence of phase transitions \cite{Habibian, Schmidt, Landig, Leonard, Piazza, Schmittberger, Baumann, Mottl}. Cold atoms offer an appealing setting to study such phenomena, since the confinement and temperature of cold gases as well as the interactions between individual atoms can be accurately controlled using external fields, which has enabled broad investigations of novel quantum phases and collective effects \cite{Baumann, Mottl, Gross, Chin}. 
For example, this includes pattern formation in Bose-Einstein condensates (BECs) of atoms with long-range dipole-dipole interactions \cite{Tanzi, Bottcher, Chomaz2019, zhang2019}, self-organization arising from light-matter interactions with cold atoms or thermal vapors \cite{Firth1990, Firth1991prl, Ackemann2001apb, Labeyire, Ackemann, Tesio, Tesio1, Firth3, Camara, Robb,Leonard, Li2017, Mivehvar,Cinti2,Sevincli, Maucher}. Generally, the rich phenomenology of these systems continues to motivate extensive research on nonlinear dynamics and self-organization  in atomic systems \cite{AckemannCommunPhys, Ackemann2018optica, Ackemann2019pra, Lutsky, Kartashov, Hou, Heinonen, Pupillo, Karpov}. One of the key elements of pattern formation are long-range interactions \cite{Mottl, Habibian, Schmidt, santos2003, Jager, Landig, Flottat, Leonard, Piazza, Ostermann}. Such interactions naturally occur between cold molecules \cite{Carr}, Rydberg atoms \cite{saffman_rev}, and dipolar quantum gases \cite{santos2003}, or can also be induced via external light fields, which offers various ways to control such photon-induced interactions via shaping and manipulating the involved photonic modes \cite{Petersen, Douglas,Mottl, Piazza, cavityRMP, Vaidya,AckemannCommunPhys, Ackemann2019pra, mirror,Boris, Matzliah}. 

\begin{figure}[t!]
\centering
  \includegraphics[width=0.8\columnwidth]{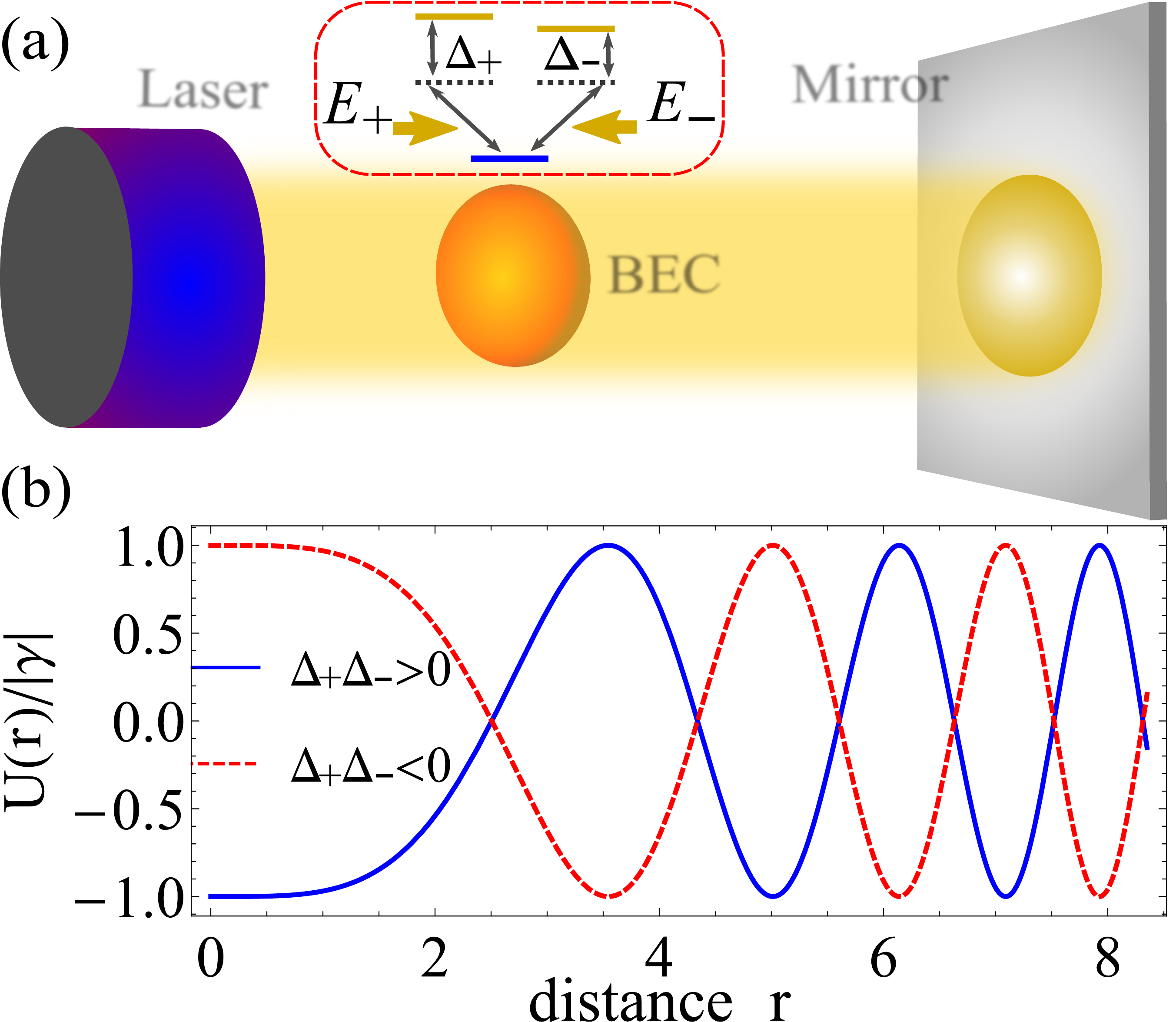}
\caption{(a) Illustration of the considered setup in which a quasi-2D BEC is illuminated by laser light that is back-reflected by a mirror and thereby generates optical feedback in the condensate. The atoms are coupled to the incoming and back-reflected light fields, with amplitudes $E_+$ and $E_-$, respectively. These light fields drive separate atomic transitions with respective frequency detunings $\Delta_+$ and $\Delta_-$, as indicated in the depicted 3-level scheme. Re-scattering of the reflected photons generates an effective atomic interaction that is shown in panel (b). The blue and red lines depict the resulting potentials for $\Delta_+\Delta_->0$ and $\Delta_+\Delta_-<0$, respectively.}
\label{fig1}
\end{figure}

Already a single mirror \cite{Firth1991prl, Labeyire, AckemannCommunPhys, Ackemann2018optica, Ackemann2019pra} can generate instabilities and structure formation through a simple optical feedback loop.  A recent work \cite{mirror} on an optically driven BEC placed in front of a retro-reflecting mirror [see Fig.\ref{fig1}(a)], showed that such a configuration leads to long-range atomic interactions that give rise to unusual patterns, such as the formation of one-dimensional chains out of two-dimensional condensates. Here, we  explore further tunability of the interaction, and in particular the control of the sign of the induced interaction via the frequencies of the driving fields, and demonstrate that a simple sign flip gives rise to a range of new states, including self-bound droplet clusters with complex lattice structures, super lattices, and supersolid states, with various rotational symmetries. Such droplet cluster states realize self-bound crystalline states that are stable without external confinement and have otherwise been elusive to matterwave settings.

The paper is organized as follows: In Sec.~\ref{Model}, we discuss the generation and tunability of photon-induced interactions via single-mirror optical feedback. The various ground states of a BEC generated by these interactions are discussed for finite systems and in the thermodynamic limit in the following sections Sec.~\ref{droplet} and Sec.~\ref{crystal}, respectively. 

\section{Photon-mediated long-range interaction}
\label{Model}

We consider a quasi-2D BEC placed in front of a retro-reflecting mirror, as shown in Fig.~\ref{fig1}(a). The atoms interact with the light field through a V-type level scheme. This coupling configuration can be realized by ensuring orthogonal polarization for the incident and reflected light field, which eliminates interference. This can be realized by using circularly polarized light and placing a quarter-wave plate in between the BEC and the mirror. As the forward light $E_+$ propagates through the condensate, it acquires a phase shift depending on the local density of the atoms, i.e., 
\begin{equation}
E_+(\mathbf{r})=E_0 e^{-i\phi({\bf r})}=E_0 e^{-i\frac{3\pi \Gamma}{2k^2\Delta_+}\rho_a({\bf r})}
\label{eq1}
\end{equation}
with $E_0$ being the amplitude of the incident light field, $\Gamma$ the spontaneous decay rate of the excited states, $k$ the wave number of the light field, $\Delta_+$ the frequency detuning between the laser light and the atomic transition. Moreover, $\rho_a=|\psi(\mathbf{r})|^2$ is the density of the condensate, and $\mathbf{r}=(x,y)$ represents the position of the atoms in the transverse plane. Following the propagation between the BEC and the mirror the back-reflected light incident on the atoms is given by 
\begin{equation}
E_-(\mathbf{r})=\frac{1}{2\pi}\int \tilde{E}_+({\bf p}) \Phi^\ast(\mathbf{p})e^{-i\mathbf{p}\cdot\mathbf{r}} {\rm d}^2 \mathbf{p}
\label{eq2}
\end{equation}
where $\tilde E_+({\bf p})$ denotes the Fourier transform of $E_+({\bf r})$, and $\Phi(\mathbf{p})=e^{2d(ik-\sqrt{\mathbf{p}^2-k^2})}$ describes the diffraction of the beam as it propagates a total distance of $2d$ towards and from the mirror, where $d$ is the distance between the BEC and the mirror as indicated in Fig.~\ref{fig1}(a).
The two light fields generate an optical dipole potential of the form 
\begin{equation}
V(\Omega_\pm)=\frac{|\Omega_+|^2}{4\Delta_+}+\frac{|\Omega_-|^2}{4\Delta_-},
\label{eq3}
\end{equation}
where $\Omega_\pm=\frac{\mu E_\pm}{\hbar}$ are the two corresponding Rabi frequencies (assuming equal dipole matrix elements $\mu$ for the two transitions) and $\Delta_-$ is the frequency detuning of $E_-$. This potential leads to an effective interaction between atoms induced by the optical feedback through the mirror. Substituting Eqs.(\ref{eq1}) and (\ref{eq2}) into Eq.(\ref{eq3}), taking the limit $|\phi({\bf r})| \ll 1$ and adiabatically eliminating the light field under conditions of large detunings, $|\Delta_{+,-}|\gg \Omega_\pm, \Gamma$, the dipole potential~(\ref{eq3}) can be rewritten as $V(\mathbf{r})=\int U(\mathbf{r}-\mathbf{r}')|\psi(\mathbf{r}')|^2{\rm d}^2 \mathbf{r}'$, with an effective atomic interaction 
\begin{equation}
U(\mathbf{r}-\mathbf{r}')=\gamma \cos{\left(\frac{|\mathbf{r}-\mathbf{r}'|^2}{4}\right)}
\label{eq4}
\end{equation}
where $\gamma=-\frac{3m\Omega^2\Gamma}{16\hbar k^2\Delta_+\Delta_-}$, $\Omega=\frac{\mu E_0}{\hbar}$, and $m$ is the atom mass. Previous work \cite{mirror} has focused on the case of identical detunings $\Delta_+=\Delta_-$, for which the interactions are attractive at short distances [i.e., $\gamma<0$, see Fig.~\ref{fig1}(b)]. This constraint can, however, be relaxed to let the detunings have opposite signs, which can for example be realized by shifting the energy of excited states with an external magnetic field \cite{Bloch1}. This allows to flip the sign of the photon-mediated interaction as illustrated in Fig.~\ref{fig1}(b). In the following, we will consider the case of $\gamma>0$ and show that, despite the oscillatory nature of the potential, this sign flip leads to profoundly different behavior of the BEC.

To this end, we consider the dynamics of the BEC as governed by the two-dimensional meanfield equation 
\begin{equation}
\begin{split}
i\frac{\partial}{\partial t}{\psi}(\mathbf{r})&=-\frac{1}{2}\nabla^2_\bot {\psi}(\mathbf{r}) +g N |{\psi} (\mathbf{r})|^2 {\psi}(\mathbf{r}) \\&+\gamma N{\psi}(\mathbf{r})\int |{\psi} (\mathbf{r}') |^2 \cos{\left(\frac{|\mathbf{r}-\mathbf{r}'|^2}{4}\right)}{\rm d}^2\mathbf{r}',
\end{split}
\label{eq5}
\end{equation}
for the condensate wave function $\psi({\bf r},t)$. Here $\nabla^2_\bot=\frac{\partial^2}{\partial x^2}+\frac{\partial^2}{\partial y^2}$, $N$ denotes the total particle number, and $g=4\pi a_s \int |\psi_z(z)|^4 dz$ is the quasi-2D strength of the contact interaction between atoms, where $a_s$ is the s-wave scattering length. The total wave function has been normalized to unity, and we have used a product ansatz ${\psi}(r)\psi_z(z)$ ($\int |\psi_z(z)|^2 dz=1$) which is valid for strong axial confinement and yields a quasi-2D description  \cite{Castin, Petrov2000}. On the other hand, there is no transverse trapping potential in the plane of the 2D condensate and all effects discussed below entirely result from the atomic interactions and do not rely on external confinement. For simplicity, space and time have been rescaled by $\sqrt{d/k}$ and $\frac{md}{\hbar k}$, respectively, to make the above equation dimensionless.

\section{Self-bound droplet clusters in small condensates }
\label{droplet}
\begin{figure}[htp]
\centering
   \includegraphics[width=\columnwidth]{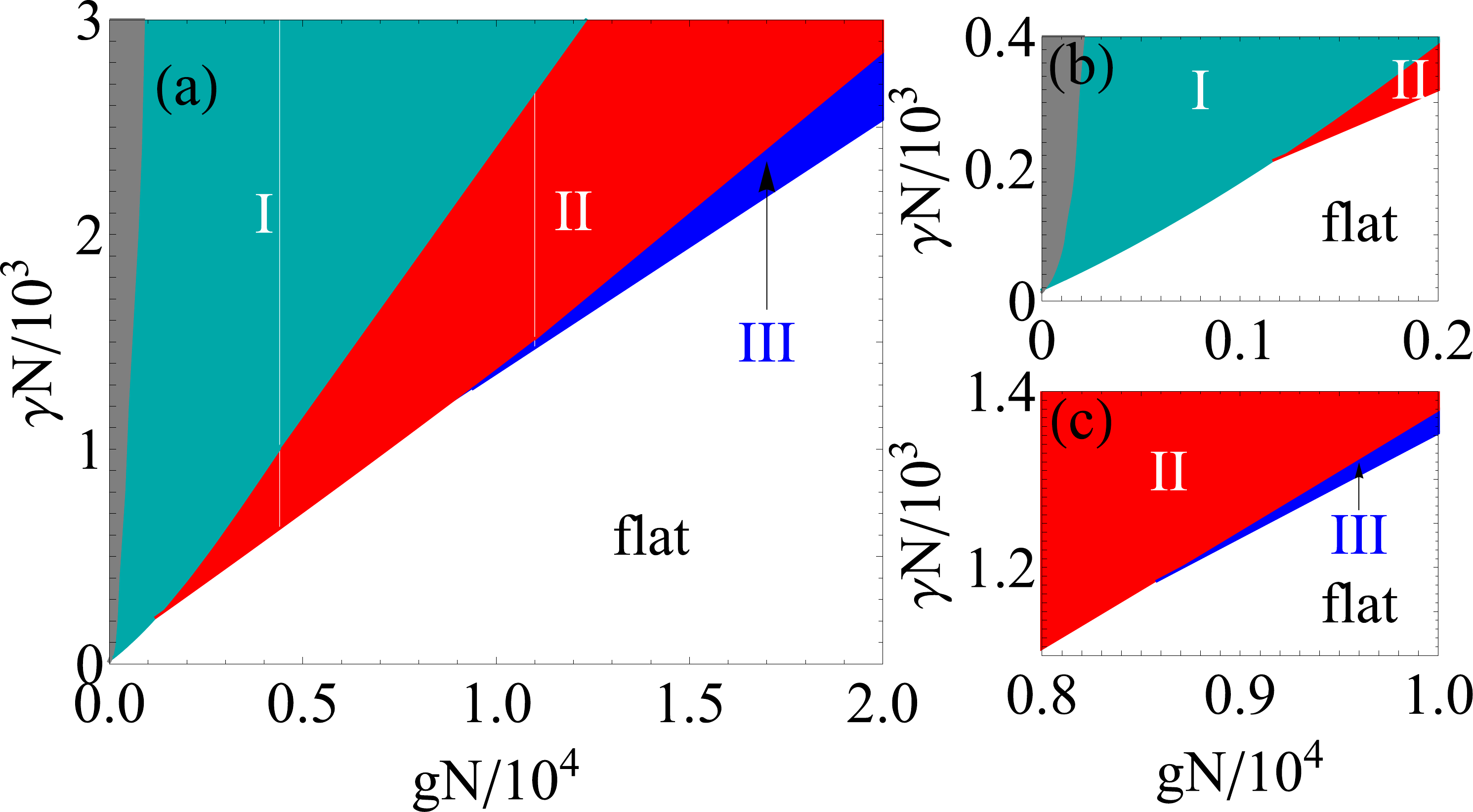}\\
   \includegraphics[width=\columnwidth]{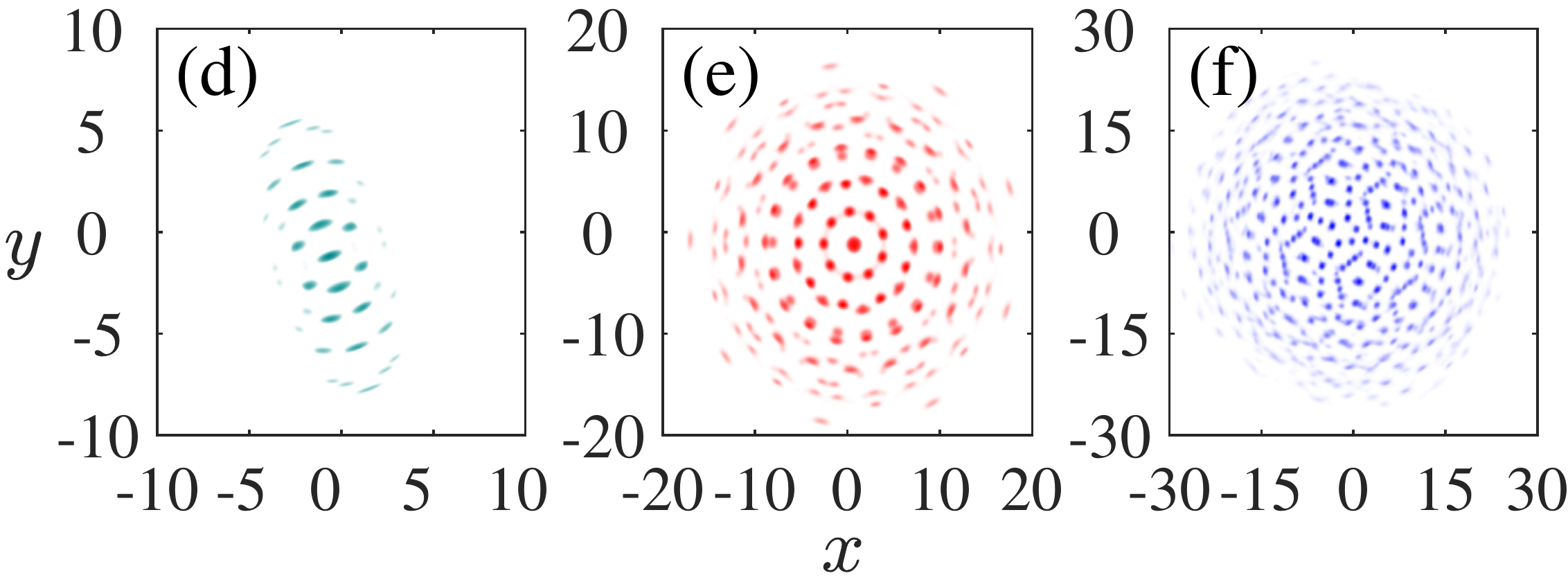}
\caption{The ground-state phase diagram of a finite BEC is displayed in (a), while (b) and (c) provide an enlarged view around the critical points. Examples of type-I ($gN/10^4=0.56$), II ($gN/10^4=1$), and III ($gN/10^4=1.2$) states are shown, respectively, in panels (d), (e), and (f) for a fixed long-range interaction strength $\gamma N/10^3=1.6$. The color depth in these plots (d), (e), and (f) represents the relative amplitude $|\psi|/|\psi|_{\rm max}$.}
\label{fig2}
\end{figure}

We explore the ground states of the nonlinear Schr\"{o}dinger equation~(\ref{eq5}) by imaginary time evolution for a fixed particle number $N$. As shown in Fig.~\ref{fig2}, the competition between zero-range and long-range interactions, as described by the two interaction strengths $g N$ and $\gamma N$, respectively, can lead to a range of different ground states.

\begin{figure}[!htp]
\centering
  \includegraphics[width=0.85\columnwidth]{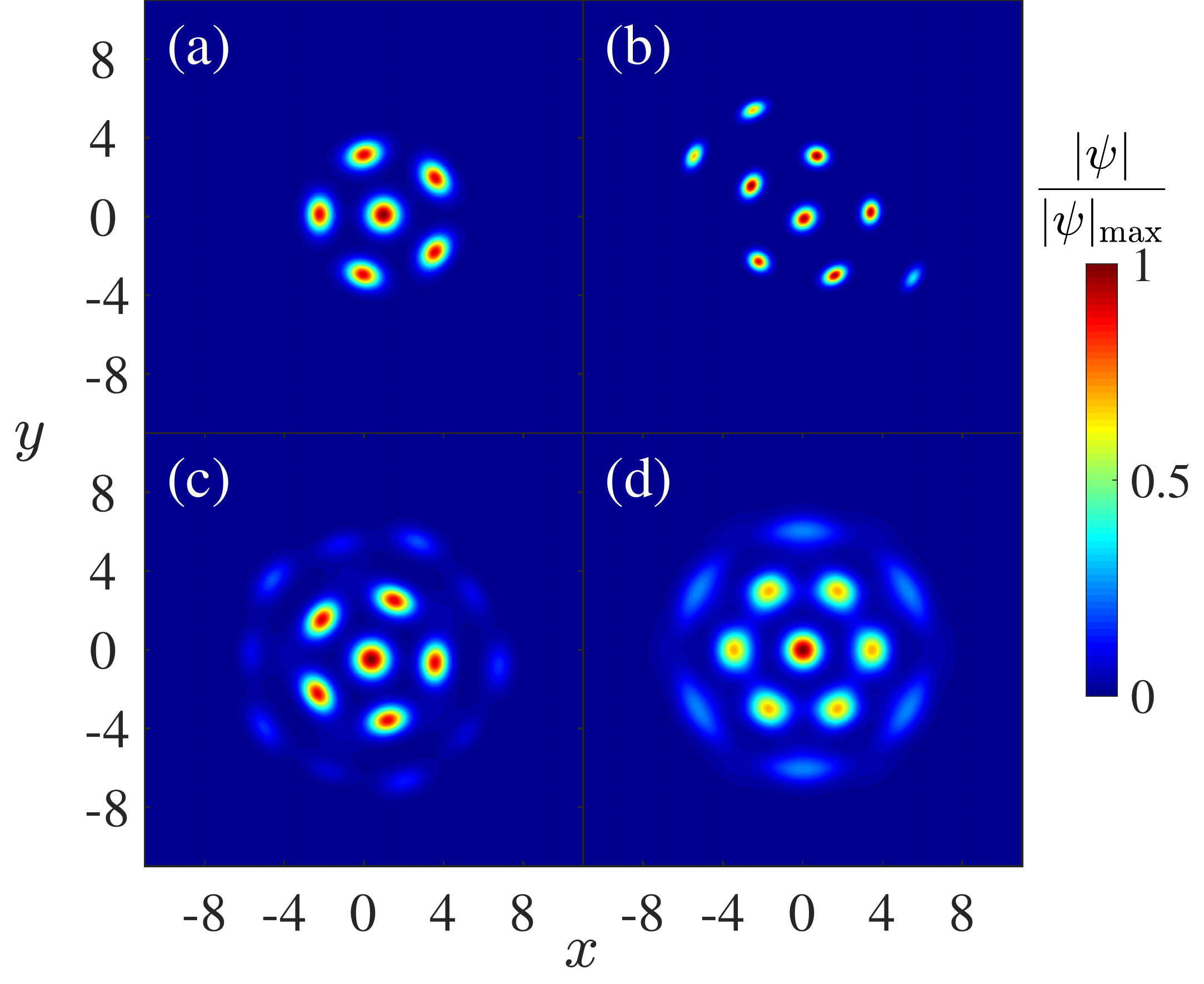}
\caption{Different pentagon-type droplet-cluster states for (a) $\gamma N=20$, $gN=0$, (b) $\gamma N=60$, $gN=0$, and (c) $\gamma N=20$, $gN=10$. A droplet cluster with 6-fold symmetry is shown in (d) at $\gamma N=20$, $gN=20$ and forms a precursor of the triangular lattice shown in Fig.\ref{fig5}(d).}
\label{fig3}
\end{figure}

We start our analysis from the regime of weak long-range interactions, where the kinetic energy plays an important role, and first discuss the case of a vanishing contact interaction. In contrast to the attractive interaction (i.e., $\gamma<0$), which tends to stabilize a single quantum droplet state \cite{mirror}, the short-range repulsion in the opposite case of $\gamma>0$ prevents the formation of single quantum droplets [see the red dashed line in Fig.~\ref{fig1}(b)]. Instead, the condensate tends to disintegrate into several small droplets. Since the sign of the long-range potential oscillates with the interatomic distance, the interaction between two droplets is attractive around the distance of $\sqrt{4(2n+1)\pi}$ ($n=0,1,2,\cdots$) and repulsive around the distance of $\sqrt{8n\pi}$ since $\gamma>0$. Consequently, the droplets bind with typical length scale of $\sqrt{4(2n+1)\pi}$, and we observe the growth of small clusters starting from small configurations into extended clusters of quantum droplets. A particular example of a small cluster with a pentagon rotational $C_5$ symmetry is displayed in Fig.~\ref{fig3}(a). Here, $C_m$ denotes the $m$-fold discrete rotational symmetry \cite{symmetry}. Note that the effective potential on the length scale of  each individual droplets is purely repulsive. Remarkably, this means that the quantum droplets are individually stabilized by their mutual attractive interaction, which also binds together the entire cluster state without external confinement. This mechanism for droplet stabilization is fundamentally distinct from that of single quantum droplets, as occurring, e.g., in two-component condensates or dipolar BECs \cite{Petrov2015, Baillie, Woachtler}. Indeed, the distances between adjacent droplets and between next-nearest neighbors are $\sim 3.5$ and $6$, respectively, which coincides exactly with the first two potential wells [cf. the minima of the red dashed line in Fig.~\ref{fig1}(b)]. This explains why the ground state in Fig.~\ref{fig3}(a) features a five-fold symmetry, that naturally exhibits these distances, rather than $C_6$ symmetry that would correspond to a triangular lattice structure, as typically occurring in 2D systems with monotonic and radially symmetric interactions. Such a six-fold symmetry can be recovered by introducing a substantial local repulsion term, as can be seen in Fig. 3(d), and apparent in the outermost shell of quantum droplets, which features larger distances between the droplets. 

The emergence of such self-bound droplet clusters is possible only beyond a critical strength $\gamma N$ of the long-range interaction. As can be seen from Fig.~\ref{fig2}(b), the condensate ground state has an unstructured density when $\gamma N<13.2$ at $gN=0$, which implies that a critical power of the driving laser is necessary to trigger the structural instability of the BEC and form droplet clusters.

\begin{figure*}[!htp]
\centering
\includegraphics[width=0.8\textwidth]{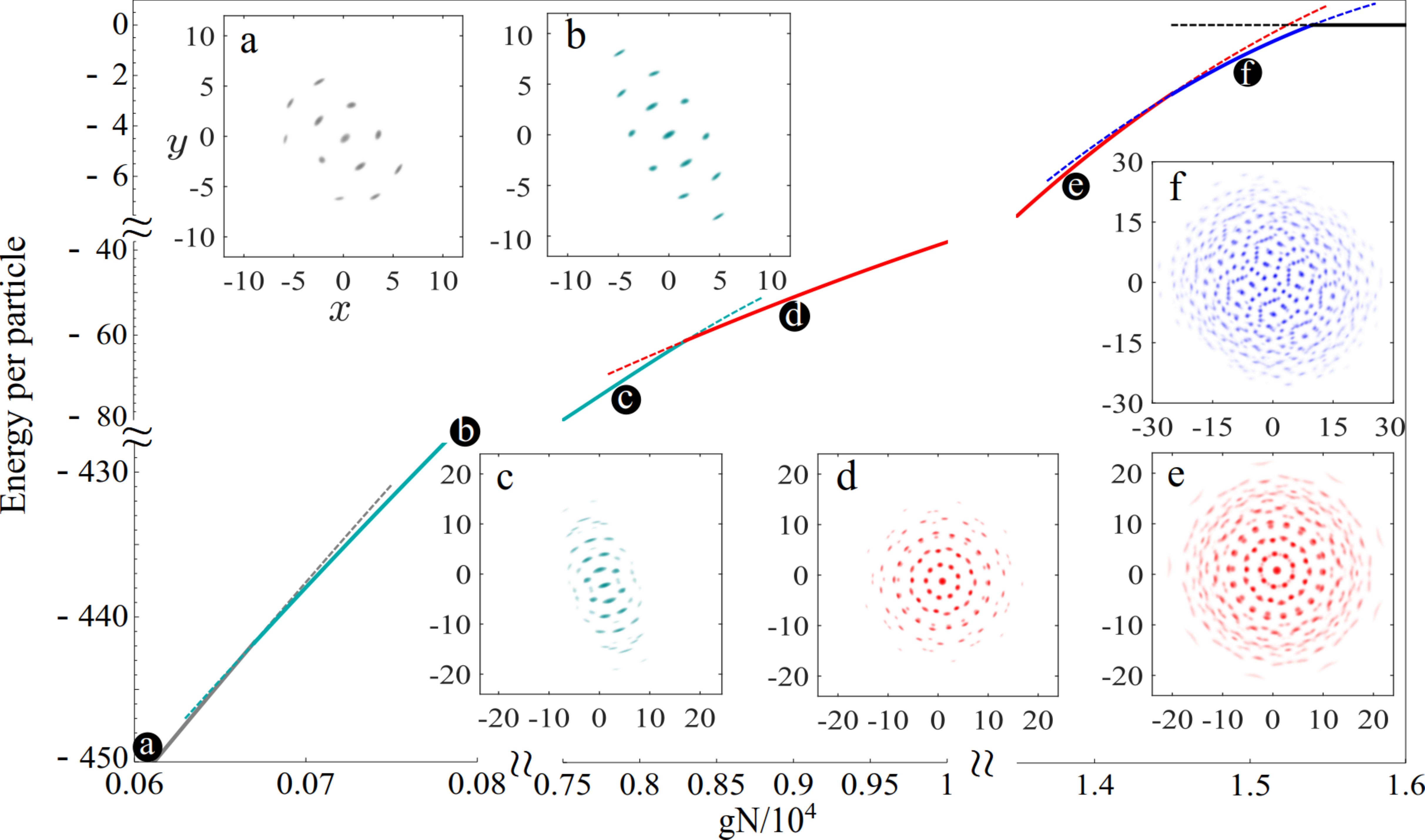}
\caption{Ground state energy as a function of the contact-interaction strength, $gN$, at a fixed strength of the long-range interaction $\gamma N=2000$. Here the solid (dashed) lines show the energy of the ground (metastable) states, and the insets display the amplitude profiles of the corresponding states at the indicated vales of $gN$. The color depth represents the relative amplitude $|\psi|/|\psi|_{\rm max}$. The energy of the trivial flat state is shown by the black line.}
\label{fig4}
\end{figure*}

By increasing the interaction strengths, $\gamma N$ and $g N$, within the gray region of Fig.~\ref{fig2}(a), the small droplet cluster shown in Fig.~\ref{fig3}(a) expand in distinct ways. First, increasing only the strength of the long-range interaction [Fig.~\ref{fig3}(b)] predominantly enhances the strength of the attractive interaction between the droplets and therefore initially tends to compress the droplets and increases the peak density of the condensate. Eventually, the correspondingly increasing positive self-interaction energy of each droplet leads to a disintegration and the formation of  separate droplets around the central pentagon, as depicted in Fig.~\ref{fig3}(b). While this can lead to the growths into cluster states without rotational symmetry [Fig.~\ref{fig3}(b)], increasing the contact interaction promotes a radial expansion of the droplet cluster and tends to restore a discrete rotational symmetry, as illustrated in Fig.~\ref{fig3}(c). Eventually, one obtains a pronounced second shell of droplet clusters, which now feature a six-fold rotational symmetry [Fig.~\ref{fig3}(d)] and are a precursor of larger and more complex droplet clusters that form at larger interactions. 

In the strong interaction regime, beyond the gray region, we can group the ground states according to their rotational symmetry into three types, illustrated in Figs.~\ref{fig2}(d)-(f). First, we find states which possess no overall rotational symmetry or merely $C_2$ corresponding to parity invariance, as illustrated in Fig.~\ref{fig2}(d) and which we label as type-I. A typical case of type-III states is shown in Fig.~\ref{fig2}(f), whereby the structure of the formed droplet cluster also features no symmetry higher than $C_2$, but which has a radially symmetric coarse-grained condensate density. Finally, there also is an extended parameter region which promotes type-II ground states possessing a high-order discrete rotational symmetry, as illustrated by the example in Fig.~\ref{fig2}(e) which features $C_9$ symmetry. Figs.~\ref{fig2}(a)-(c) show the transitions between the different BEC ground states. The corresponding ground state energies are shown in Fig.~\ref{fig4} as a function of the contact interaction strength $gN$ and demonstrate that all found transitions are of first-order type, with a finite region of bistability of the two competing orders around each structural transition.

\section{Droplet crystals in the thermodynamic limit}
\label{crystal}

In the previous section, we discussed the ground states in the case of a small BEC of finite particle number, and demonstrated the emergence of novel self-bound droplet cluster states with rather peculiar structures. Now, we can take the thermodynamic limit by considering $N\rightarrow\infty$ at a fixed average density $\rho$ of the atoms.

\begin{figure}[!htp]
\centering
  \includegraphics[width=0.9\columnwidth]{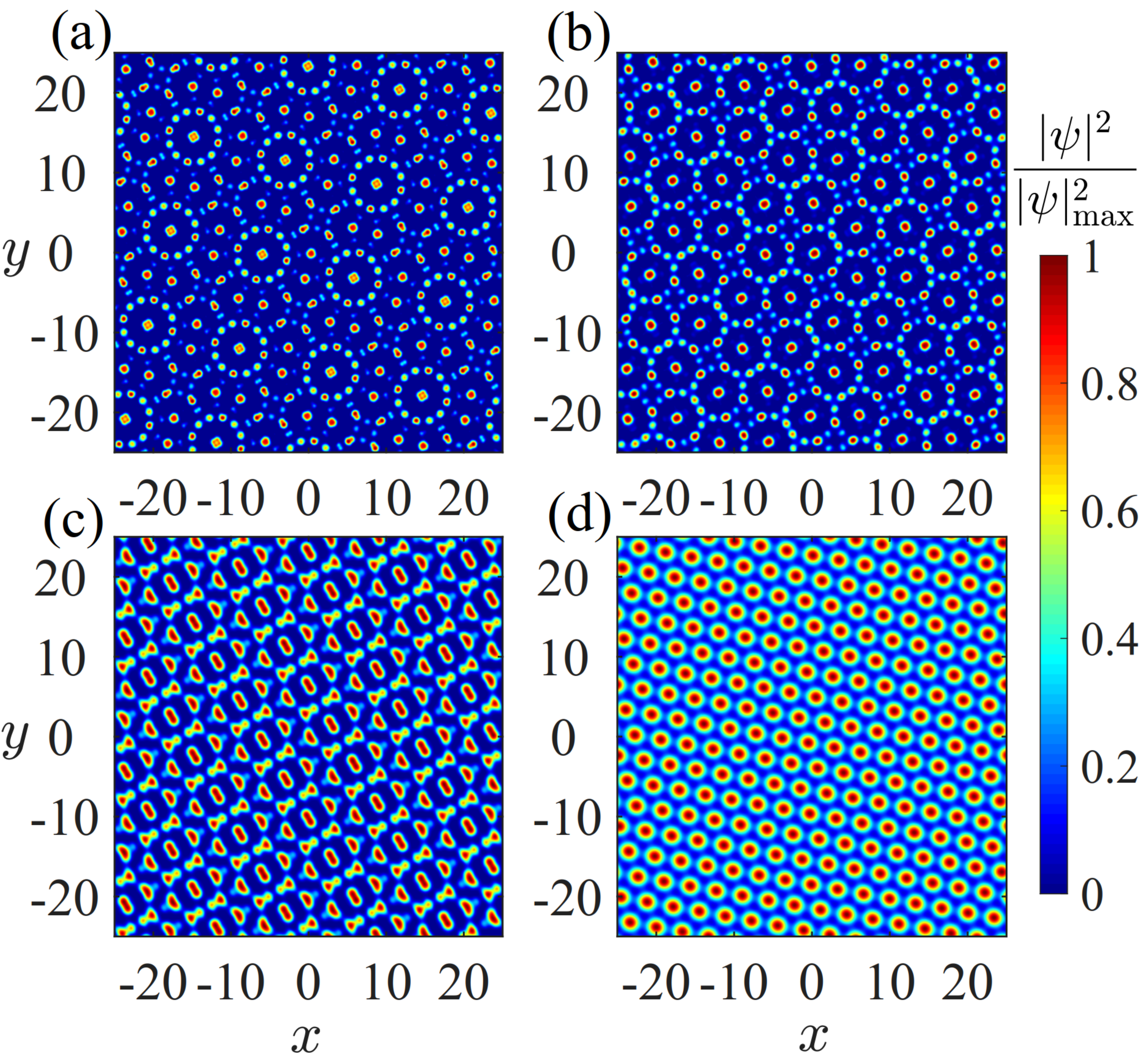}
\caption{Exemplary, density profiles of extended cluster crystals  with (a) a 4-fold rotational symmetry ($g\rho = 28$), (b) with no rotational symmetry ($g\rho = 31$), (c) with a two-fold rotational symmetry ($g\rho = 34$), and (d) with a 6-fold rotational symmetry ($g\rho=36$). The strength of the long-range interaction is fixed at $\gamma \rho = 3$.}
\label{fig5}
\end{figure}

This leads to the formation of extended droplet crystals which, once again, have a remarkably rich and complex structure as illustrated in Fig.~\ref{fig5}. Indeed, the structure of the ground state changes dramatically with increasing contact interaction at fixed long-range interaction, which is kept at $\gamma \rho=3$ in Fig.~\ref{fig5}. Such crystals with superlattice structures have also attracted interest in different contexts \cite{Hellwig, Fontecha, Cole, Stoica}. The present case provides a simple formation mechanism, whereby the fine carpet-like structures arise from the oscillatory behavior of the interaction and the distance-dependent period of these oscillations. It is, therefore, absent in the more common case of monotonic interactions \cite{Sevincli, Maucher, Cinti2, Cinti1, Hsueh, pfau1}. The characteristic superlattice structures can be controlled by the competition between the interaction strengths $g$ and $\gamma$, whereby the large-$g$ limit recovers the more common case of simple triangular lattices as shown in Fig.~\ref{fig5}(d). In addition, we can again identify three types of droplet crystals for weaker contact interactions. The first type of crystals shown in Fig.~\ref{fig5}(a) is composed of smaller lattice-site structures that each feature a discrete rotational symmetry similar to the finite droplet clusters shown in Fig.~\ref{fig2}(e). As the contact interaction is increased, another kind of lattice structure emerges which is now formed by two types of basic units that do not necessarily feature rotational symmetry. Despite the fairly complex nature of the formed density pattern, depicted in Fig.~\ref{fig5}(b), it nevertheless is fully periodic, i.e. features discrete translational symmetry. At even larger contact interactions, a third type of crystal state appears, as shown in Fig.~\ref{fig5}(c). Here, the unit cell of this rather unusual structure has no rotational symmetry (higher than $C_2$) but has an approximately hexagonal shape, akin to the compressed hexagonal phases observed for classical cluster crystals with Lifshitz-Petrich-Gaussian interaction potentials \cite{LPG}.

\begin{figure}[!htp]
\centering
  \includegraphics[width=0.8\columnwidth]{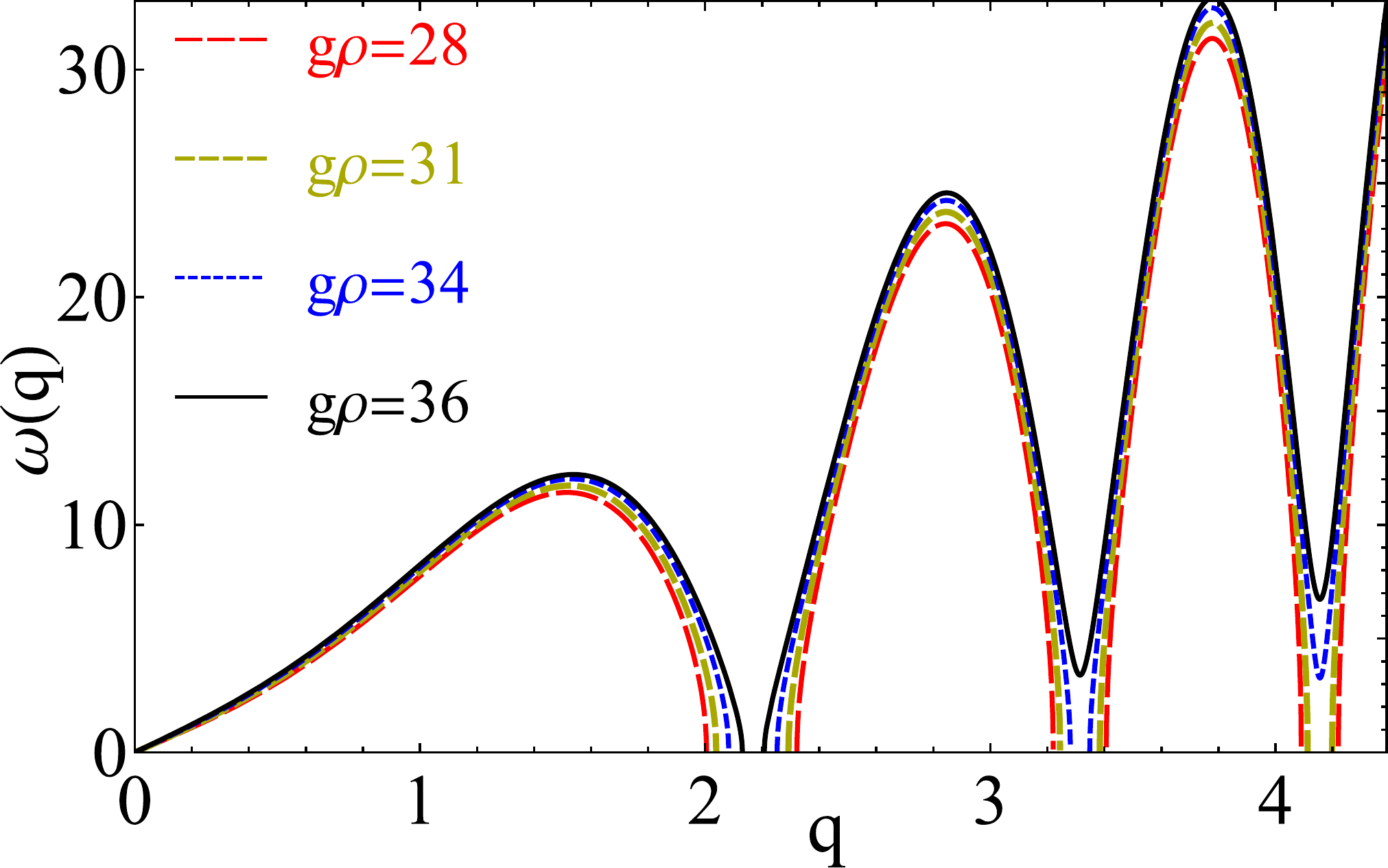}
\caption{Excitation spectrum of the condensate for different strengths, $g\rho$, of the contact interaction at a fixed long-range interaction strength $\gamma \rho = 3$.}
\label{fig6}
\end{figure}

To gain more intuitive insights into the origins of the diverse crystal structures, we consider the excitation spectrum of a homogeneous BEC. Following standard Bogoliubov theory \cite{Bogoliubov, Ozeri}, the energy $\omega$ at a given momentum $q$ of the excitation reads
\begin{equation}
\omega(q)=\sqrt{\frac{q^2}{2}\left(\frac{q^2}{2}+2\rho \left(g +4\pi \gamma  \sin{q^2}\right)\right)},
\label{eq6}
\end{equation}
and is shown for different values of the interaction strength in Fig.~\ref{fig6}. As can be seen, the spectrum features multiple roton minima at momenta $q_{\rm rot}=\sqrt{(2n+3/2)\pi}$ ($n=0,1,2,\cdots$). For sufficiently weak contact interactions, several of these roton modes soften simultaneously which gives rise to several distinct length scales for supported density waves. It is the competition between these length scales that leads to the complex crystal structures depicted in Fig.~\ref{fig5}(a)-(c). Indeed, as we increase the contact interaction to the point where only the roton mode with the lowest momentum softens (black line in Fig.~\ref{fig6}), one recovers the simple triangular lattice structure of Fig.~\ref{fig5}(d). Note that close to the first roton instability, the amplitude of the triangular density wave decreases, leaving a substantial background density between the otherwise isolated quantum droplets. This implies a finite global superfluidity of the system \cite{Leggett, Chester}, and therefore the formation of a 2D supersolid state, in which superfluidity and crystalline order coexist. 

\section{Conclusion}
\label{con}
In summary, we studied the generation of long-range atomic interactions via a single optical feedback loop in a BEC, and have described a three-level driving scheme that permits to controlling the sign of the light-induced interactions. Changing the interactions from attractive to repulsive gives rise to a remarkably rich spectrum of distinct ground states in the form of mesoscopic clusters and extended crystals of quantum droplets. Importantly, these structures are entirely self-bound, in stark contrast to regular patterns found condensates with monotonic atomic interaction potentials. The found clusters and crystals can have diverse symmetry properties that can be controlled, e.g., by the atomic density. These include extended droplet crystals, found in the thermodynamic limit, which can not only have a 6-fold for simple triangular lattices, but also two-fold or four-fold rotational symmetry, or regular crystals with no rotational symmetry at all. Within a simplified picture, the emergence of such diverse ground states can be traced back to the excitation spectrum of the homogeneous BEC, which features multiple roton minima of comparable energy and which can soften simultaneously to generate competing length scales in the system. The dynamics of the underlying roton instability can be studied in experiments by rapidly turning on the light-induced interaction while the described ground states could be prepared via an adiabatic increase of the laser intensity to slowly turn on the long-range interaction between the atoms. Here, photon emission from the weakly populated excited states presents the main limitation of available timescales for observing the effects of light-induced interactions. While off-resonant laser driving suppresses spontaneous emission and can provide sufficient time to observe pattern formation under typical experimental condition \cite{Labeyire, AckemannCommunPhys, Ackemann2018optica, Ackemann2019pra}, this limitation may be further alleviated via optical resonators, i.e. using multiple mirrors instead of a single-mirror feedback loop. Further perspectives derive from the flexibility of the mechanism to generate interactions. For example, driving different hyperfine states offers a simple approach to creating quantum-gas mixtures with competing long-range interactions, whereby, e.g., the coexistence of attractive and repulsive interactions can lead to diminished mean-field effect, and thus the demonstrated control of the sign of the interaction could be used to enhance and study effects of quantum fluctuations \cite{Petrov2015}.

\section{Acknowledgement}
We thank Thorsten Ackemann and Fabian Maucher for fruitful discussions and helpful comments. This work has been supported by the EU through the H2020-FETOPEN Grant No. 800942640378 (ErBeStA), by the DFG through the SPP1929, by the Carlsberg Foundation through the Semper Ardens Research Project QCooL, by the DNRF through a Niels Bohr Professorship to T. P. and the DNRF Center of Excellence CCQ (Grant Agreement No. DNRF156), and by the NSF through a grant for the Institute for Theoretical Atomic, Molecular, and Optical Physics at Harvard University and the Smithsonian Astrophysical Observatory.

\end{document}